\documentclass[referee]{raa}            

\usepackage{graphicx,times}             
\input{epsf.sty}                  

\begin{document}
\
\title{Multi-wavelength study of triggered star formation around 25 H {\large II} regions
}

   \volnopage{Vol.0 (200x) No.0, 000--000}      
   \setcounter{page}{1}          

   \author{Jin-Long  Xu
      \inst{1,2}
   \and Jun-Jie Wang
      \inst{1,2}
      \and Chang-Chun Ning
      \inst{2,3}
      \and Chuan-Peng Zhang
      \inst{1,2,4}
   }

   \institute{National Astronomical Observatories, Chinese Academy of Sciences,
             Beijing 100012, China {\it xujl@bao.ac.cn}\\
        \and
             NAOC-TU Joint Center for Astrophysics, Lhasa 850000, China\\
             \and
             Tibet University, Lhasa,  Tibet, 850000, China\\
             \and
             University of the Chinese Academy of Sciences, Beijing, 100080,  China\\
   }

\date{Received~~2013 month day; accepted~~2013~~month day}

\abstract{We have investigated 25 H {\small II} regions with bubble morphology in $^{13}$CO(1-0) and infrared data, to search the quantitative evidence for the triggering star formation by  collect and collapse (CC) and radiatively driven implosion (RDI) models. These H {\small II} regions display the morphology of the complete or partial bubble at 8 $\mu$m, which all are associated with the surrounding molecular clouds. We obtained that the electron temperature ranges from 5627 K to 6839 K in these H {\small II} regions, and the averaged electron temperature is 6083 K. The age of these H {\small II} regions is from 3.0$\times10^{5}$ yr to 1.7$\times10^{6}$ yr, and the mean age is 7.7$\times10^{5}$ yr.
Based on the morphology of the associated molecular clouds, we divided these H {\small II} regions into three groups, which may support CC and RDI models. We selected 23 young $IRAS$ sources with the infrared luminosity of $>$10$^{3}$$L_{\odot}$ in 19 H {\small II} regions. In addition, we identified some young stellar objects (including Class I sources), which are concentrated only in H {\small II}
regions G29.007+0.076, G44.339-0.827, and G47.028+0.232. The polycyclic aromatic hydrocarbons (PAHs) emission of the three H {\small II} regions all show the cometary globule. Comparing the age of each H {\small II} region with the characteristic timescales for star formation, we
suggest that the three H {\small II} regions can trigger the clustered star formation by RDI process. In addition, we for the first time detected seven molecular outflows in the five H {\small II} regions. These outflow sources may be triggered by the corresponding H {\small II} regions.
   }

   \authorrunning{J.-L. Xu et al. }            
   \titlerunning{Multi-wavelength study of triggered star formation around 25 H {\small II} regions}  

   \maketitle

%
%
\section{Introduction}           
\label{sect:intro} Isolated low-mass star formation has been well understood, the
scenario of massive star formation and clustered star formation remain unclear.
The feedback of massive star has significant impact on the surrounding environment,
such as  H {\small II} region, stellar winds, and supernova
remnant (SNR)(\cite{Zinnecker07,Dewangan12}). On the one hand, when these energetic products of massive star advance into the surrounding molecular clouds,
it causes overdensities along the advancing boundary and some dense cores form. On the other hand,  when these energetic products of massive star collide with some pre-existing compact cloud cores,  it will trigger the cloud cores to collapse. The triggered stars may form more quickly than quiescent isolated star formation.
Moreover, since H {\small II} region, stellar winds, and SNR may impact on the large range of molecular clouds,
then the clustered star formation may be triggered. Previous observations toward some SNRs (\cite{Junkes92,Paron09,Xu11,Xu12}) suggest that
although some identified young stars are distributed around the SNRs, the obtained age of these SNRs is not enough large than time scales of young stars.
What did trigger these young stars?  SNRs may explode in the molecular shell produced by the H {\small II} region and/or stellar winds of their progenitor (\cite{Junkes92,Paron09,Su09,Zhou09}), then the young stars found in the surrounding of SNRs may be triggered by the H {\small II} region and/or stellar winds.

Previous studies toward individual H {\small II} regions suggest that H {\small II} regions can trigger the young star formation (for example,\cite{Zavagno10,Dewangan12,Paron11,Dirienzo12}).
Two processes have been considered for the triggering of star formation at the edge of the H {\small II} regions, namely `collect and collapse (CC)'(\cite{Dale07}) and `radiation
driven implosion (RDI)' (\cite{Sandford82}). In the CC process, a compressed layer of high-density neutral material forms between the ionization front and shock front preceding it in the neutral gas, and star formation occurs when this layer becomes gravitationally unstable, first proposed by Elmegreen \& Lada (1997). Massive star formation may be triggered by CC process (\cite{Whitworth94}); In RDI process, the shocks drive into pre-existing density structures and compress them to form stars, which is characterized by the cometary globules or optically bright rims. Bright-rimmed clouds (BRCs) found in H {\small II} regions are potential sites of triggered star formation by RDI process. The RDI process may trigger the low-mass and intermediate-mass star formation (\cite{Reipurth83}).
Recently, observational searches for evidence of triggered star formation focused on \cite{Churchwell06} bubbles. \cite{Churchwell06} concluded that about 25\% of the bubbles coincide with known radio H {\small II} regions, while Deharveng et al. (2010) suggested that about 86\% of the bubbles contain ionized gas detected by the radio-continuum emission.  Although nearly all GLIMPSE bubbles may be caused by H {\small II} regions, about half of all Galactic H {\small II} regions have a bubble morphology in the infrared band (\cite{Bania10}).

In this paper we will study 25 H {\small II} regions with bubble-like morphologies in a same way to analyze triggered star formation in H {\small II} regions. In Section 2, we summarize the
selection of sample and young stars. In Section 3, we give the general
results and discussion.  In
Section 4, we summarize our main conclusions.

\section{DATA}
\subsection{Sample selection}
The target sample was constructed by applying the following criteria: (1) The selected H {\small II} regions are in the GBT-Arecibo-GRS-NVSS-GLIMPSE surveys overlap regions.
(2) The size of each selected H {\small II} region must be greater than
the beam size of the surveys. (3) The selected H {\small II} regions have a bubble morphology at 8 $\mu$m. From this criteria, we selected 25 H {\small II} regions as our sample, 16 of which are associated with  \cite{Churchwell06} bubbles. Table 1 lists the target sample source parameters including the source name, the Galactic longitude and latitude, the LSR velocity of H$_{\alpha}$, and integrated flux density (S). The properties of these surveys are summarized below.

In the GBT survey (\cite{Anderson111}), 448 previously unknown Galactic H {\small II} regions were detected at X-band (9 GHz, 3 cm) in the Galactic zone 343$^{\circ}$$\leq$$\ell$$\leq$67$^{\circ}$ and $|b|$$\leq$1$^{\circ}$. The FWHM beam size
of the telescope is approximately 82$^{\prime\prime}$ at this frequency. Additionally, in the Arecibo H {\small II} region surveys, \cite{Bania12} reported the discovery of 37 previously unknown H {\small II} regions with the Arecibo telescope at X-band (9 GHz, 3 cm). The Galactic zone of survey is 31$^{\circ}$$\leq$$\ell$$\leq$66$^{\circ}$ with $|b|$$\leq$1$^{\circ}$. Its beam is nearly three times smaller than that of the GBT.

To trace the ionized gas and investigate the molecular gas distribution of the surrounding H {\small II} regions, we used the  NVSS and BU-FCRAO GRS surveys. NVSS is a 1.4 GHz continuum survey
covering the entire sky north of -40$^{\circ}$ declination (\cite{Condon98})  with a noise of about 0.45 mJy/beam and the resolution of 45$^{\prime\prime}$. While the GRS is a new
survey of Galactic $^{13}$CO(1-0) emission (\cite{Jackson06}). The survey covers a longitude range of $\ell$$=$18$^{\circ}$--55.7$^{\circ}$ and a latitude range of $|b|$$\leq$1$^{\circ}$, with a  angular resolution of 46$^{\prime\prime}$. The survey's velocity coverage is -5 to 135 km s$^{-1}$ for Galactic
longitudes $\ell$$\leq$40$^{\circ}$ and -5 to 85 km s$^{-1}$ for Galactic longitudes $\ell$$>$40$^{\circ}$. At the velocity resolution of
0.21 km s$^{-1}$, the typical rms sensitivity is 0.13 K.

GLIMPSE survey is used to identify the young stars along H {\small II} regions, which observed the Galactic plane (65$^{\circ}$
$<$ $|l|$ $<$ 10$^{\circ}$ for $|b|$ $<$ 1$^{\circ}$) with the four
mid-IR bands (3.6, 4.5, 5.8, and 8.0 $\mu$m) of the Infrared Array
Camera (IRAC) (\cite{Fazio04}) on the Spitzer Space
Telescope. The resolution is from 1.5$^{\prime\prime}$ (3.6 $\mu$m) to
1.9$^{\prime\prime}$ (8.0 $\mu$m).

\begin{figure}
\vspace{0mm}
\includegraphics[angle=0,scale=0.95]{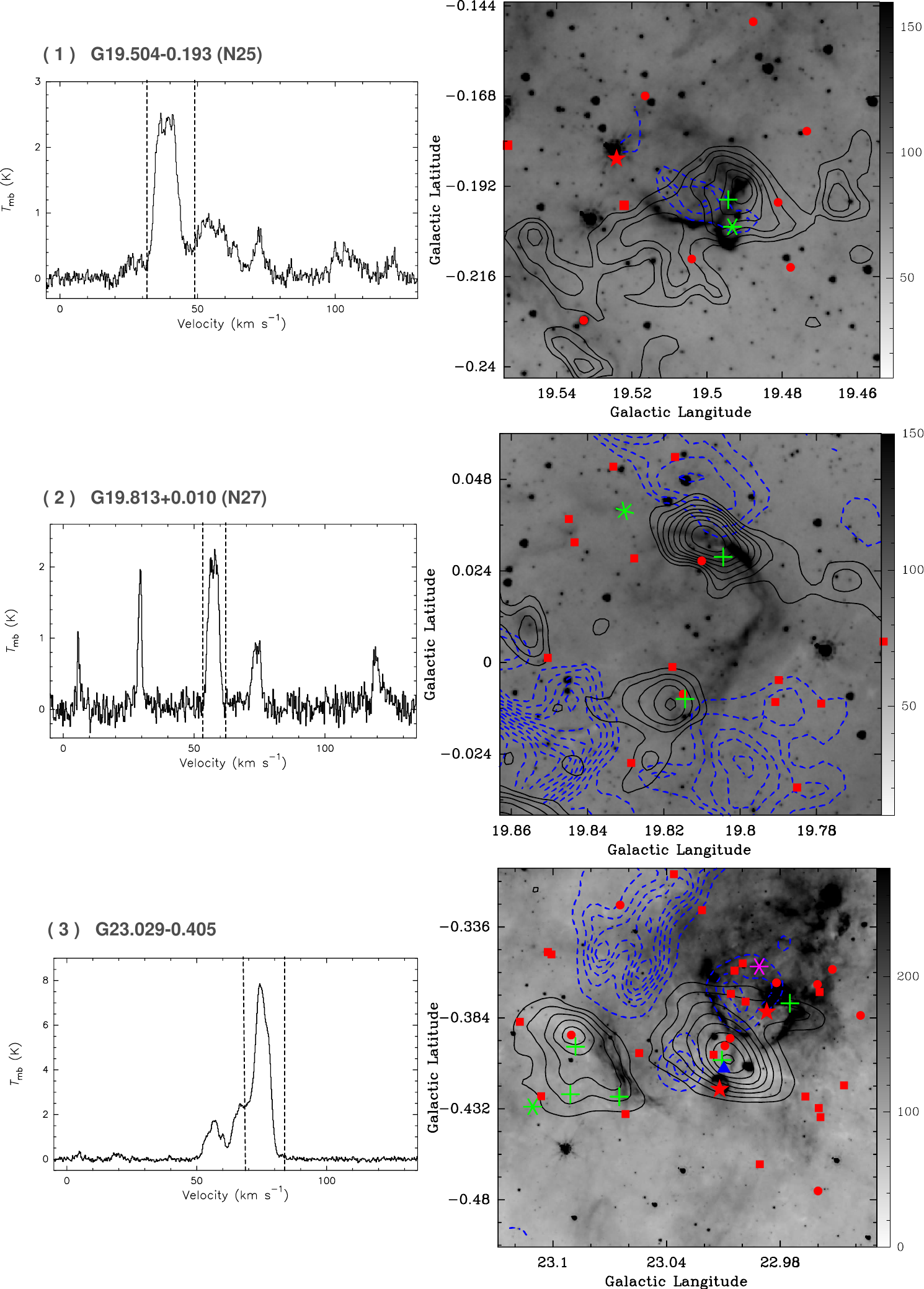}
\vspace{-5mm}\caption{Left panels: $^{13}$CO J=1-0
spectral profile averaged over the molecular clouds. Two black dashed lines and red solid line
mark the integrated velocity ranges and tangent point velocity, respectively. Right panels: $^{13}$CO J=1-0 integrated
intensity (black contours) and 1.4 GHz continuum emission (blue dashed contours) maps overlayed on the mid-infrared 8
$\mu$m emission map (grey scale). The b
lue contour levels begin at the peak flux of 20$\%$ and increase in steps of the peak flux of 10$\%$. Class I sources are labeled as the
red dots, and Class II sources as the red fulled square. The red stars represent IRAS sources . The green pluses and $``\ast"$  indicate the millmeter continuum sources and X ray sources, respectively. The pink $``\ast"$ marks the supernova remnant candidate. The contour levels are (1) 10.0, 10.7, 11.4, 12.1, 12.8, 13.6, 14.3 km s$^{-1}$. (2) 2.9, 3.4, 3.8, 4.3, 4.7, 5.2, 5.6, 6.1 km s$^{-1}$. (3) 19.3, 22.0, 24.5, 27.5, 30.2, 32.9, 35.6, 38.3 km s$^{-1}$. }
\end{figure}

\begin{figure}
\vspace{0mm}
\includegraphics[angle=0,scale=.95]{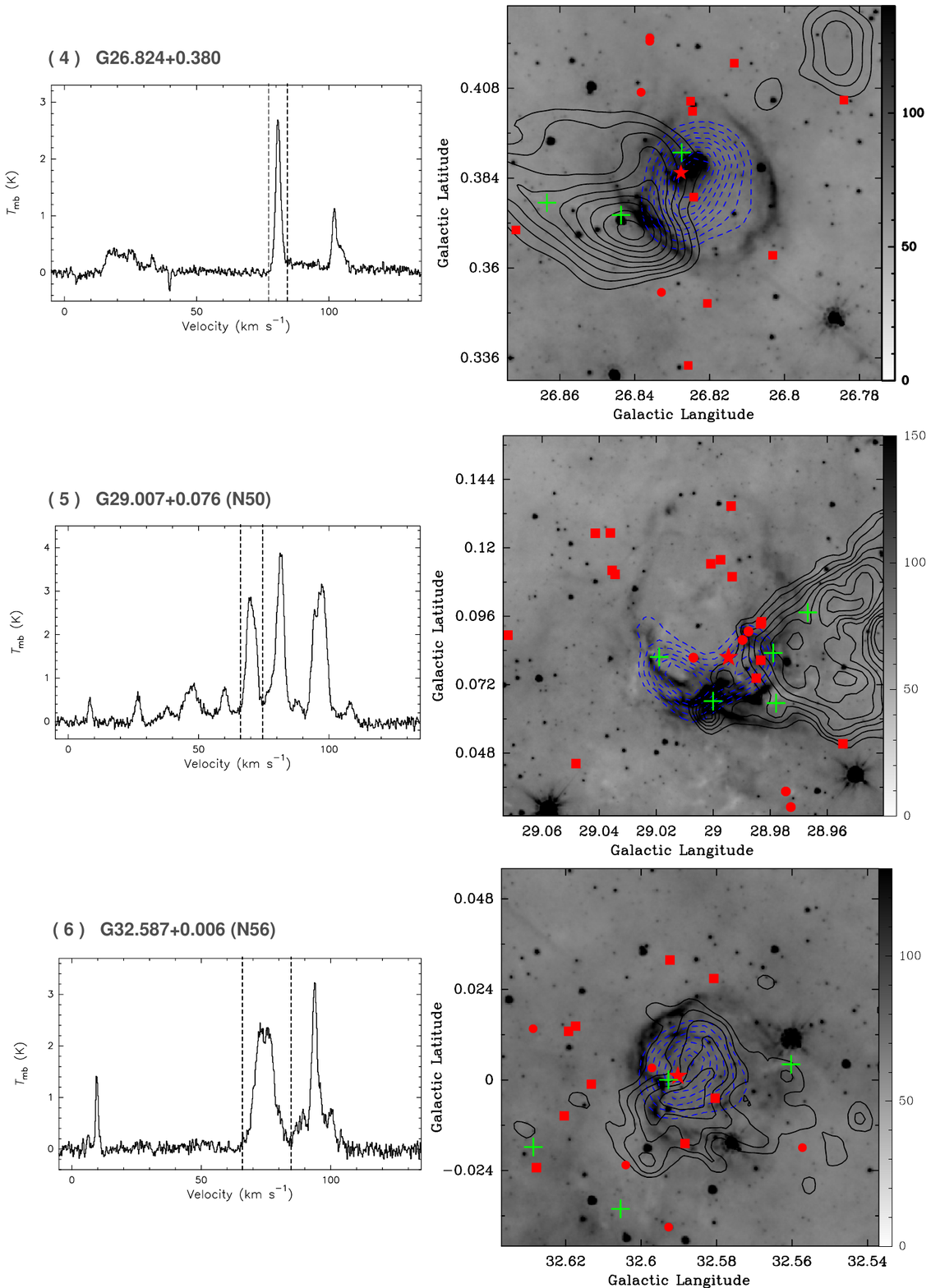}
{Continue: (4) 2.0, 2.4, 2.7, 3.1, 3.5, 3.8, 4.2, 4.5, 4.9 km s$^{-1}$. (5) 3.5, 4.0, 4.7, 5.4, 6.0, 6.7, 7.4, 8.0 km s$^{-1}$. (6) 5.2, 6.3, 7.3, 8.3, 9.4, 10.4, 12.5, 13.6, 14.6 km s$^{-1}$. (7) 5.2, 6.3, 7.3, 8.3, 9.4, 11.5, 12.5, 13.6, 14.6 km s$^{-1}$. (8) 2.5, 2.8, 3.1, 3.5, 3.8, 4.2, 4.5, 4.8 km s$^{-1}$. (9) 3.7, 4.3, 4.8, 5.4, 6.0, 6.5, 7.1, 7.7 km s$^{-1}$. (10) 5.9, 6.5, 7.1, 7.7, 8.3, 8.9, 9.5, 10.1, 10.7, 11.3, 11.8 km s$^{-1}$. (11) 5.3, 6.1, 6.9, 7.7, 8.5, 9.3, 10.2, 11.0 km s$^{-1}$. (12) 2.7, 3.3, 3.8, 4.3, 4.9, 5.4, 6.0, 6.5, 7.0, 7.6 km s$^{-1}$. (13) 7.6, 9.1, 10.7, 12.2, 13.7, 15.2, 16.7, 18.3, 19.8, 21.3 km s$^{-1}$. (14) 2.7, 3.2, 3.8, 4.3, 4.9, 5.4, 5.9, 6.5, 7.0, 7.5 km s$^{-1}$. (15) 4.2, 5.2, 6.2, 7.2, 8.2, 9.2, 10.2, 11.2 km s$^{-1}$. (16) 2.3, 3.0, 3.6, 4.3, 4.9, 5.6, 6.2, 6.9, 7.5, 8.2, 8.8 km s$^{-1}$. (17+18) 0.9, 1.3, 1.8, 2.2, 2.7 km s$^{-1}$. (19) 4.6, 5.3, 5.9, 6.6, 7.3, 8.0, 8.7, 9.4, 10.1, 10.7, 11 km s$^{-1}$.  (20) 7.8, 8.8, 9.7, 10.6, 11.6, 12.5, 13.4, 14.4, 15.3 km s$^{-1}$. (21) 1.6, 2.1, 2.6, 3.2, 3.7, 4.2, 4.7, 5.3 km s$^{-1}$. (21) 4.7, 5.9, 7.1, 8.4, 9.6, 10.9, 11.1, 13.3, 14.6 km s$^{-1}$. (23) 0.8, 1.4, 2.0, 2.6, 3.2, 3.8, 4.4, 5.0, 5.6, 6.2, 6.8, 7.4 km s$^{-1}$. (24) 2.9, 3.4, 3.9, 4.4, 4.9, 5.5, 6.0, 6.5, 7.0 km s$^{-1}$. (25) 4.1, 4.9, 5.6, 6.3, 7.0, 7.8, 8.5, 9.2, 9.9 km s$^{-1}$. }
\end{figure}

\begin{figure}
\vspace{0mm}
\centering
\includegraphics[angle=0,scale=1.35]{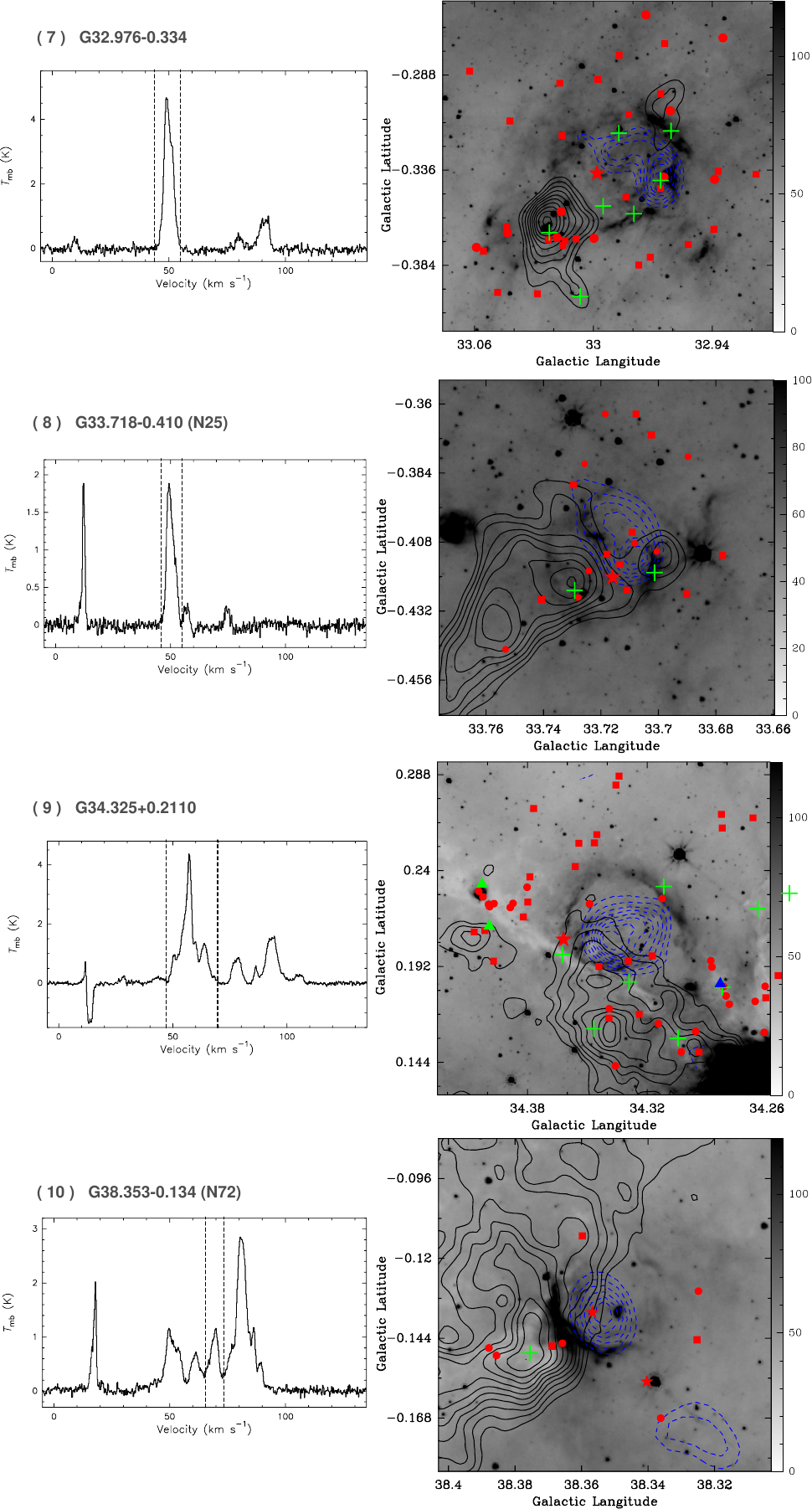}
\vspace{15mm}
\end{figure}

\begin{figure}
\vspace{0mm}
\centering
\includegraphics[angle=0,scale=1.35]{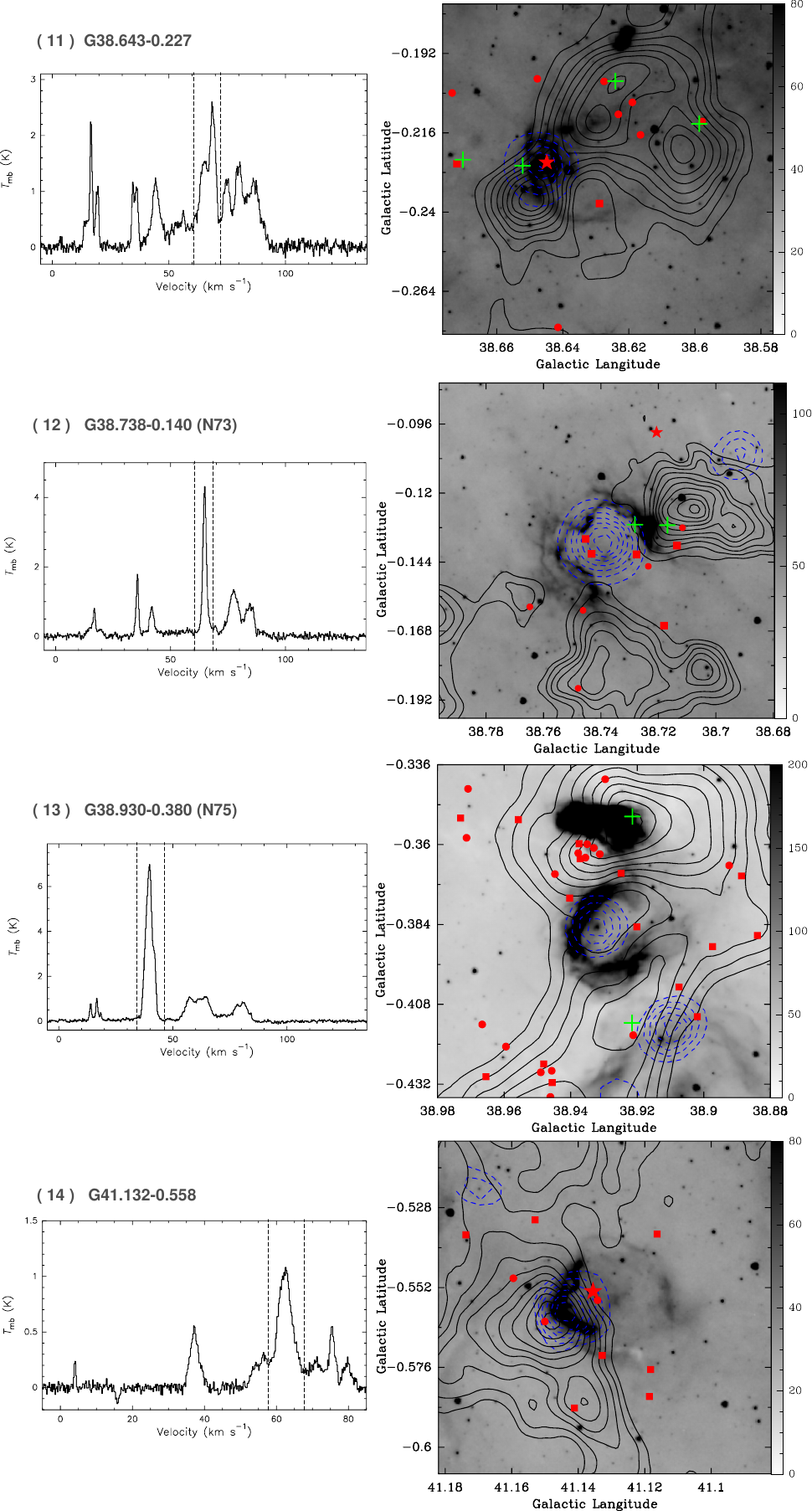}
\vspace{15mm}
\end{figure}

\begin{figure}
\vspace{0mm}
\centering
\includegraphics[angle=0,scale=1.35]{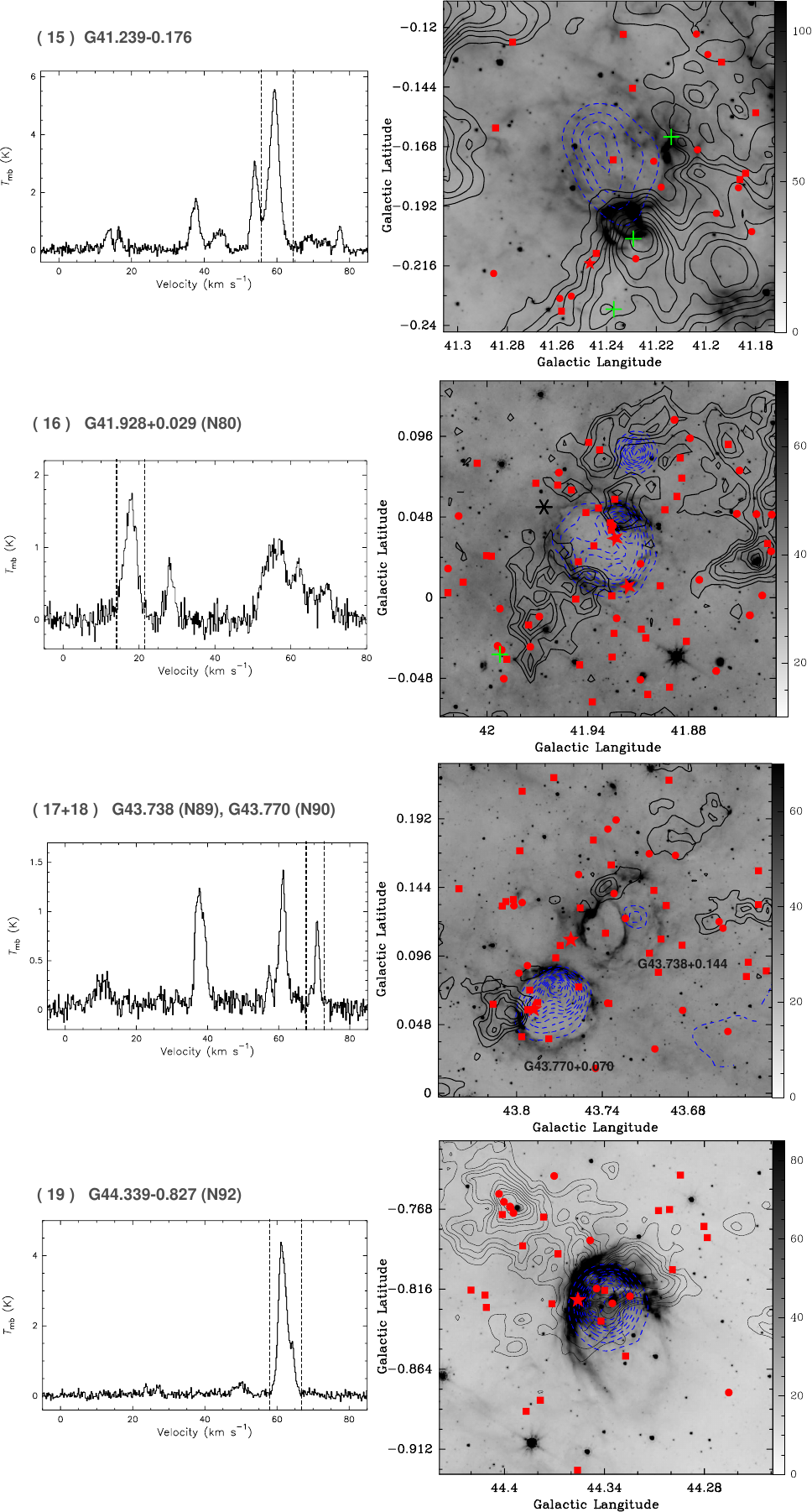}
\vspace{15mm}
\end{figure}

\begin{figure}
\vspace{0mm}
\centering
\includegraphics[angle=0,scale=1.35]{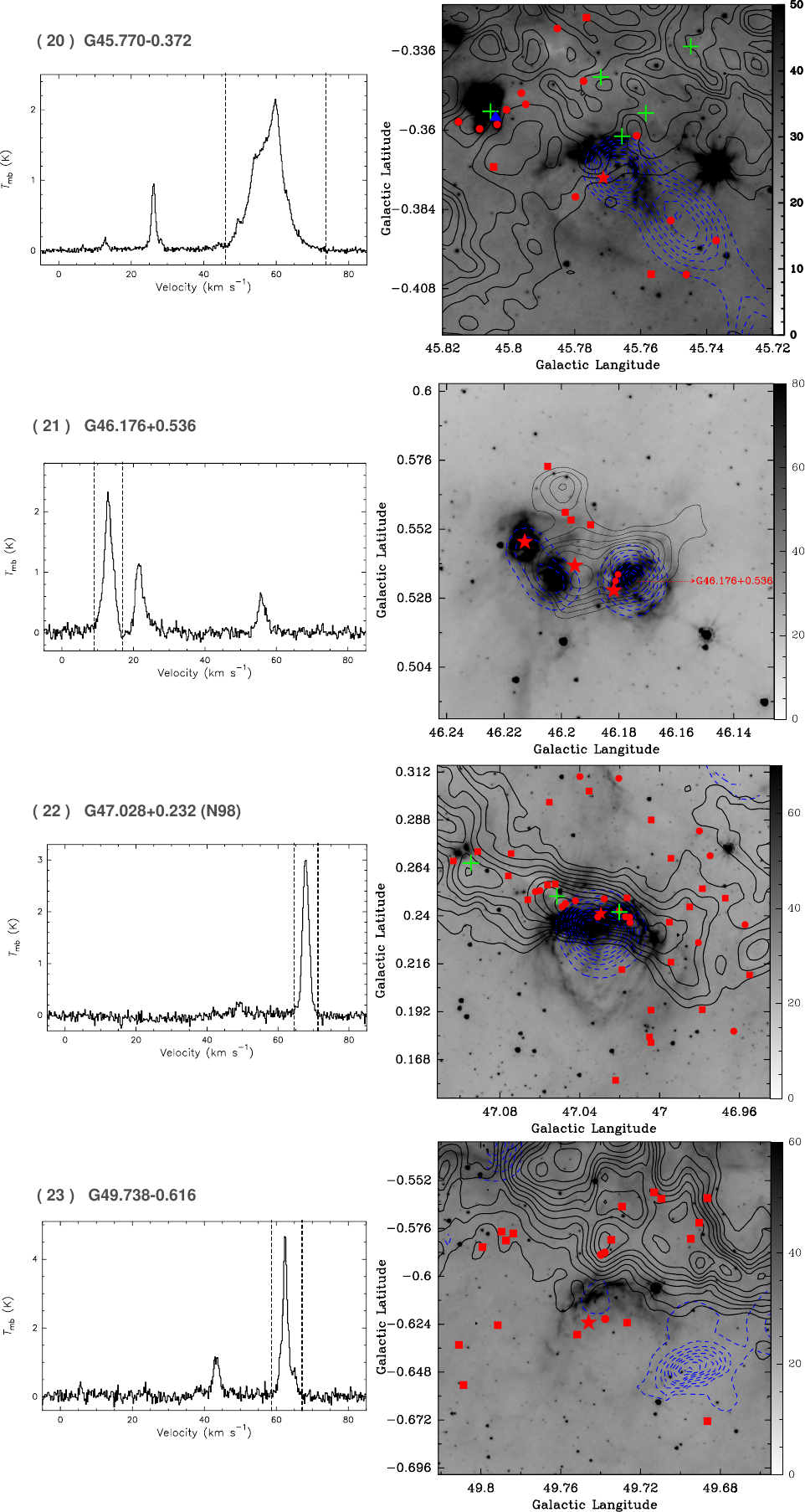}
\vspace{15mm}
\end{figure}

\begin{figure}
\vspace{0mm}
\centering
\includegraphics[angle=0,scale=1.15]{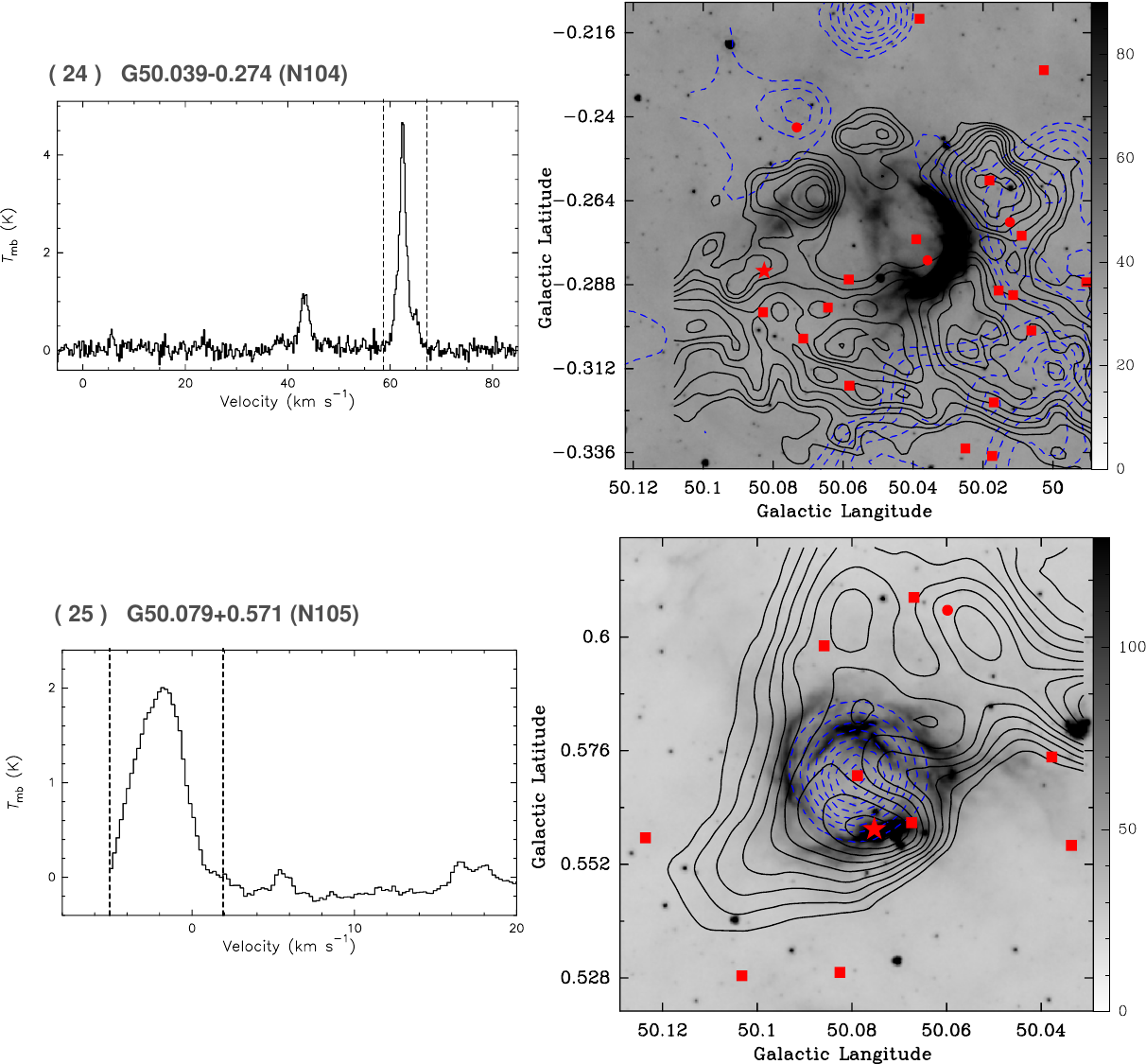}
\vspace{5mm}
\end{figure}

\section{Results and Discussion}
\label{sect:data}
\subsection{H {\small II} regions}
\subsubsection{The morphology of H {\small II} regions}
Figure 1 (Right panels) shows the 1.4 GHz continuum emission contours (blue) superimposed on  Spitzer-IRAC 8 $\mu$m image of each H {\small II} region (color scale). The 8 $\mu$m IRAC band contains emission from polycyclic aromatic hydrocarbons (PAHs). PAHs are believed to be destroyed in the ionized gas, but thought to be excited at the interface of H {\small II} region and molecular cloud by the radiation leaking from the H {\small II} region. Hence, the triggered formation of young stars are easily found in the molecular clouds adjacent to 8 $\mu$m emission. In Figure 1 (Right panels), the 8 $\mu$m emission display the morphology of the complete and partial bubble.
The 1.4 GHz continuum emission reveals the presence of ionized gas in each H {\small II} region. From Figure 1 (Right panels), we can see that the 8 $\mu$m bubble of these H {\small II} regions are filled with the 1.4 GHz continuum emission, except for G19.813+0.010 and 50.039-0.274. Hence, PAH emission may be excited by the radiation leaking from these H {\small II} regions. In G41.928+0.029 region, there may be another H {\small II} region located at the northwest, which is surrounding by the shell-like molecular gas. Through the morphology at 8 $\mu$m, we measure the sizes of each bubble, which are listed in Table 1.

\subsubsection{The distance of H {\small II} regions}
CO data are widely used to trace the morphology of molecular clouds. The $^{13}$CO(1-0) average spectrum in each H {\small II} region are presented in Figure 1 (Left panels). From these panels, we can see that spectrum of twenty-two H {\small II} regions show multiple velocity components detected along the line of sight, except for H {\small II} regions G44.339-0.827, G47.028+0.232, and G49.738-0.616. According to the hydrogen radio recombination line (RRL) velocity components of each H {\small II} region given by \cite{Anderson11}, we select the velocity components of $^{13}$CO(1-0) associated with the H {\small II} regions, whose peak is located between two vertical dashed lines in Figure 1 (Left panels).  Vis
Gaussian fits to all the spectra,  we obtain the central
line velocity  ($V_{\rm LSR}$), the peak intensity($T_{\rm mb}$), and full width at half-maximum
(FWHM).

\begin{table}
\begin{center}
\tabcolsep 0.4mm\caption{H {\small II} region sample and CO
parameters}
\begin{tabular}{lccccccccccccccccc}
\hline\hline
 Number &Region & $l$ & $b$ &$T_{\rm mb}$ & FWHM & $V_{\rm LSR}$ & $V_{\rm H_{\alpha}}$ & S  & D & R& $N_{\rm H_{2}}$ &$n_{\rm i}$&\\
 &    & deg & deg  & (K)  &  (km $\rm s^{-1}$)  &  (km $\rm s^{-1}$)  &  (km $\rm s^{-1}$) & mJy & kpc &pc&  cm$^{-2}$ & $10^{3}$cm$^{-2}$ &\\
\hline
1&G19.504-0.193  & 19.504 & -0.193& 2.6 &9.4(0.4)  & 38.9(0.2) & 37.8 & 168 & 12.8$^{a}$   & 2.8 &1.5$\times10^{22}$&3.5& N25\\
2&G19.813+0.010  & 19.813 & +0.010& 2.3 &4.4(0.2)  & 57.5(0.1) & 60.4 & 187 & 11.8$^{b}$  & 5.8 &6.3$\times10^{21}$&1.7& N27\\
3&G23.029-0.405  & 23.029 & -0.405& 7.4 &7.7(0.2)  & 74.6(0.1) & 76.0 & 1014& 4.7$^{a}$      & 3.6 &3.6$\times10^{22}$&3.2&  \\
4&G26.824+0.380  & 26.824 & +0.380& 2.6 &2.4(0.1)  & 80.9(0.1) & 82.0 & 120 & 4.9$^{a}$    & 2.2 &3.9$\times10^{21}$&1.5& N44 \\
5&G29.007+0.076  & 29.007 & +0.076& 2.8 &4.8(0.6)  & 70.3(0.2) & 67.7 & 454 & 11.5$^{a}$   & 7.1 &8.4$\times10^{21}$&1.2& N50\\
6&G32.587+0.006  & 32.587 & +0.006& 2.4 &8.4(0.4)  & 74.5(0.2) & 77.4 & 273 &  9.7$^{a}$    & 4.3 &1.3$\times10^{22}$&4.1& N56\\
7&G32.976-0.334  & 32.976 & -0.334& 4.5 &4.4(0.1)  & 49.8(0.1) & 49.3 & 492 &11.0$^{a}$   & 7.3 &1.2$\times10^{22}$&2.5&  \\
8&G33.718-0.410  & 33.718 & -0.410& 1.8 &3.9(0.2)  & 49.9(0.1) & 53.3 & 151 &3.3$^{a}$    & 1.4 &4.4$\times10^{21}$&2.7&  \\
9&G34.325+0.211 & 34.325 & +0.211& 3.9 &4.0(0.4)   &56.9(0.1) & 62.9  & 307 &10.3$^{a}$   & 5.9 &9.8$\times10^{21}$&1.2& N62\\
10&G38.353-0.134 & 38.353 & -0.134& 1.2 &4.0 (0.5) & 70.3(0.2) & 70.5 & 93.0$^{d}$   &8.6$^{b}$  & 3.2 &3.0$\times10^{21}$&0.5& N72\\
11&G38.643-0.227 & 38.643 & -0.227& 2.6 &5.1(0.3)  & 67.9(0.3) & 63.8 & 86  &8.7$^{a}$    & 3.1 &8.3$\times10^{21}$&2.4&  \\
12&G38.738-0.140 & 38.738 & -0.140& 4.1 &2.0(0.1)  & 64.9(0.1) & 60.9 & 213 &8.9$^{a}$    & 3.2 &5.1$\times10^{21}$&1.1& N73\\
13&G38.930-0.386 & 38.930 & -0.386& 6.6 &3.9(0.1)  & 39.7(0.1) & 42.1 & 38  &2.7$^{a}$    & 0.7 &1.6$\times10^{22}$&6.2& N75\\
14&G41.132-0.558 & 41.132 & -0.558& 2.5 &3.8(0.4)  & 62.4(0.1) & 65.5 & 14.2$^{d}$   &6.4$^{c}$  & 2.5 &5.9$\times10^{21}$&1.7& \\
15&G41.239-0.176 & 41.239 & -0.176& 6.1 &10.0(0.3) & 59.0(0.5) & 56.9 & 440  &8.7$^{a}$    & 3.8 &3.8$\times10^{22}$&9.0&  \\
16&G41.928+0.029 & 41.928 & +0.029& 1.6 &4.0(0.4)  & 17.7(0.2) & 20.7 & 620  &11.4$^{a}$   & 5.7 &4.0$\times10^{21}$&0.8& N80\\
17&G43.738+0.114 & 43.738 & +0.114& 1.8 &1.6(0.3) & 71.0(0.1) & 73.1 & 15.4$^{d}$   &6.1$^{c}$  & 4.2 &2.3$\times10^{21}$&0.6& N89\\
18&G43.770+0.070 & 43.770 & +0.070& 1.2 &2.1(0.2) & 68.7(0.1) & 70.5 & 360 &6.1$^{a}$     & 3.8 &2.3$\times10^{21}$&1.2& N90  \\
19&G44.339-0.827 & 44.339 & -0.827& 4.0 &3.3(0.1)  & 61.7(0.1) & 62.5 & 136  &6.1$^{a}$    & 4.1 &8.3$\times10^{21}$&1.7& N92\\
20&G45.770-0.372 & 45.770 & -0.372& 1.8 &10.0(0.2) & 58.0(0.1) & 51.0 & 15.0$^{d}$   &5.8$^{b}$  & 2.2 &1.1$\times10^{22}$&0.8&  \\
21&G46.176+0.536 & 46.176 & +0.536& 2.1 &3.0(0.2)  & 13.0(0.1) & 6.3  & 30.0$^{d}$   &10.8$^{c}$   & 2.5 &3.9$\times10^{21}$&0.9&  \\
22&G47.028+0.232 & 47.028 & +0.232& 4.3 &3.7(0.1)  & 56.4(0.1) & 56.9 & 390  &5.8$^{a}$  & 3.8 &1.0$\times10^{22}$&1.6& N98\\
23&G49.738-0.616 & 49.738 & -0.616& 3.0 &2.2(0.1)  & 67.8(0.1) & 62.4 & 37.3$^{d}$   &5.5$^{c}$  & 3.8 &4.1$\times10^{21}$&0.8&  \\
24&G50.039-0.274 & 50.039 & -0.274& 4.1 &2.0(0.2)  & 62.5(0.1) & 60.9 & ---   &5.5$^{c}$  & 2.9 &5.1$\times10^{21}$&1.9& N104\\
25&G50.079+0.571 & 50.079 & +0.571& 2.1 &3.3(0.1)  & -2.0(0.1) & -1.1 & 195  &11.1$^{a}$ & 3.8 &4.1$\times10^{21}$&0.5& N105\\
\hline\hline
\end{tabular}
\end{center}
Notes. $^{a}$Distances are determined with the H {\small I} E/A and $^{13}$CO(1-0) method.
$^{b}$Distances are only based on the judgement of $^{13}$CO(1-0) about the kinematic distance ambiguity.
$^{c}$Distances are obtained from  the judgement of \cite{Bania12} about the kinematic distance ambiguity. $^{d}$The integrated flux at 1.4 GHz.
\end{table}

Based on the Galactic
rotation model of \cite{Fich} together with
$R_{\odot}$ = 8.5 kpc and $V_{\odot}$ = 220 km s$^{-1}$, where
$V_{\odot}$ is the circular rotation speed of the Galaxy, we derive
the kinematic distances of these H {\small II} regions. However, an H {\small II} region at a specific radial velocity can be at either the near or far kinematic distance, resulting in a distance ambiguity. The H {\small I} self-absorption features product when cool H {\small I} clouds absorb the warmer H {\small I} emission from the background H {\small II} region continuum emission. H {\small I} emission/absorption (H {\small I} E/A) is considered as an effective method to distinguish the distance ambiguity (\cite{Kuchar94,Kolpak03,Anderson111}). \cite{Anderson11} and \cite{Bania12} have resolved the kinematic distance ambiguity of twenty-two our selected H {\small II} regions with this method. Because it is difficult to identify the H {\small I} absorption associated with the H {\small II} region above error estimates, they did not give the distances of another four H {\small II} regions. For H {\small I} E/A method, whether every sight line in between the near and far distance existing cool H {\small I} clouds is of great importance.  The cool H {\small I} clouds also are traced by optical thin $^{13}$CO(1-0), whose emission is relatively stronger. H {\small I} E/A, with the addition of $^{13}$CO(1-0),  may accurately resolve the kinematic distance ambiguity. If $^{13}$CO(1-0) emission coincidents with H {\small I} absorption is detected between the H {\small II} region velocity and the tangent-point velocity, it suggests an H {\small II} region at the far distance. On the contrary, it is at near distance.  Using H {\small I} E/A and $^{13}$CO(1-0) method, we check the distance of 13 H {\small II} regions from GBT H {\small II} regions survey, and give the distances of G26.824+0.380, G33.718-0.410, G38.930-0.386, and G41.239+0.029. Because \cite{Bania12} did not give the H {\small I} absorption spectrum of seven H {\small II} regions from Arecibo H {\small II} regions survey, we yet adopted their judgement about the kinematic distance ambiguity. In addition, we also can not give the judgement of another three H {\small II} regions G19.813+0.010, G38.353-0.134, and G45.770-0.372. In 23 of 25 selected H {\small II} regions, 12 are at the far distance, 8 are at the tangent-point distance, and  3 are at the near distance. In Figure 1 (Left panels), we find that if an H {\small II} region is far distance, $^{13}$CO(1-0) spectrum of the associated molecular clouds has several velocity components in between the tangent-point and system velocity. Because  G19.813+0.010 and G38.353-0.134 have several velocity components in between the tangent-point and system velocity, we suggest that both H {\small II} regions may be at far distance. The velocity of G45.770-0.372 is near tangent-point velocity, we adopt the tangent-point distance as that of G45.770-0.372. The fitted and obtained results are summarized in Table 1.

\subsubsection{The age of H {\small II} regions}
Above analysis suggest that the characteristics of the surrounding medium may affect the processes of triggered star formation. Moreover, the time scale is also important for deciding  the triggered star formation. Assuming each H {\small II} region expanding in a homogeneous medium, the dynamical ages of each H {\small II} region can be estimated by using the model (\cite{Dyson80})
\begin{equation}
\mathit{t_{\rm HII}}=7.2\times10^{4}(\frac{R_{\rm H{\small II}}}{\rm pc})^{4/3}(\frac{Q_{\rm Ly}}{10^{49}s^{-1}})^{-1/4}(\frac{n_{\rm i}}{10^{3}\rm cm^{-3}})^{-1/2} \rm yr,
\end{equation}\indent
where $R_{\rm HII}$ is the radius of the H {\small II} region, $n_{\rm i}$ is the initial number density of the gas and $Q_{\rm Ly}$ is the ionizing luminosity. Because the 8 $\mu$m bubble of these H {\small II} regions are filled with the 1.4 GHz continuum emission, we take the size of the bubbles  as the radius of the H {\small II} regions. In previous studies toward several H {\small II} regions ( \cite{Zavagno06}; \cite{Deharveng08}; \cite{Paron09,Paron11}; \cite{Pomares09}; \cite{Dirienzo12}), they all determined an initial number density of $\sim$10$^{3}$cm$^{-3}$. Here, we also adopt an initial number density of $\sim$10$^{3} $cm$^{-3}$ for all the H {\small II} regions. Assuming the emission is optically thin free-free thermal continuum, $Q_{\rm Ly}$ was computed by \cite{Mezger74}
\begin{eqnarray}
 \mathit{Q_{\rm Ly}}=4.761\times10^{48}a(\nu,T_{e})^{-1}(\frac{\nu}{\rm GHz})^{0.1}(\frac{T_{e}}{\rm K})^{-0.45}(\frac{S_{\nu}}{\rm Jy})\times(\frac{D}{\rm kpc})^{2}\rm~photons~s^{-1},
\end{eqnarray}

Where $\nu$ is the frequency of the observation, $S_{\nu}$ is the observed specific flux density, and $D$ is the distance to the H {\small II} region. $a(\nu,T_{e})$ is defined by \cite{Mezger67}
\begin{eqnarray}
a(\nu,T_{e})=0.366(\frac{T_{e}}{\rm K})^{-0.15}(\frac{\nu}{\rm GHz})^{0.1}\{\rm ln[4.995\times10^{-2}
\times(\frac{\nu}{\rm GHz})^{-1}]
+1.5\rm ln(\frac{\it T_{e}}{\rm K})\}
\end{eqnarray}

$T_{e}$ is the electron temperature given by \cite{Paladini04} and \cite{Tibbs12}
\begin{equation}
\mathit{T_{e}}=(4166\pm124)+(314\pm20)\times d~\rm K,
\end{equation}\indent
According to the rotation curve models of \cite{Fich}, the Galactocentric distance of H {\small II} regions $d$ can be given by
 \begin{equation}
 \mathit{d}=\frac{V-a_{1}V_{\odot}}{a_{2}\omega_{\odot}},
\end{equation}\indent
Where $V$ is the rotation velocity around the Galactic center of an H {\small II} region  at a Galactocentric distance, and $\omega_{\odot}$ is the rotation velocity and angular velocity of the Sun. The fitting coefficients $a_{1}$, $a_{2}$ and $\omega_{\odot}$ are 0.99334, 0.0030385 and 27.5 km s$^{-1}$ (\cite{Fich89}), respectively. In the Arecibo H {\small II} region survey, because they did not give the flux density of each H {\small II} region at 9 GHz, we replace the flux by that at 1.4 GHz. The obtained electron temperature, ionizing luminosity, and age of 18  H {\small II} regions are listed in Table 2. From Table 2, we see that $T_{e}$ ranges from 5627 K to 6839 K in these H {\small II} regions, and the averaged $T_{e}$ is 6083 K; $a$($\nu$,$T_{e}$) is about 0.9 for all the H {\small II} regions. Moreover, $t$$_{_{\rm H{\small II} }}$  ranges from 3.0$\times10^{5}$ yr to 1.7$\times10^{6}$ yr, and the mean age is 7.7$\times10^{5}$ yr.

\begin{table*}
\begin{center}
\tabcolsep 1.4mm\caption{The obtained parameters of H {\small II} regions
}
\begin{tabular}{cccccccccccccccc}
\hline\hline
 Number &Region  &$d$ & $T_{e}$ & a($\nu$,$T_{e}$) & $Q_{\rm Ly}$ & $t_{_{\rm H{\small II}}}$ & \\
 &      & (kpc)  &  K  &    &  photons~$\rm s^{-1}$ & yr &\\
\hline
1&G19.504-0.193  & 5.6 &5935$\pm$237  & 0.9 & (3.7$\pm$0.4)$\times10^{48}$ & (3.6$\pm$0.1)$\times10^{5}$  \\
2&G19.813+0.010  & 4.9 &5694$\pm$221  & 0.9 & (3.5$\pm$0.5)$\times10^{48}$ & (9.8$\pm$0.4)$\times10^{5}$  \\
3&G23.029-0.405  & 4.7 &5627$\pm$217  & 0.9 & (3.1$\pm$0.4)$\times10^{48}$ & (5.3$\pm$0.2)$\times10^{5}$  \\
4&G26.824+0.380  & 4.8 &5673$\pm$220  & 0.9 & (3.9$\pm$0.7)$\times10^{47}$ & (4.7$\pm$0.2)$\times10^{5}$  \\
5&G29.007+0.076  & 5.2 &5802$\pm$228  & 0.9 & (8.1$\pm$0.7)$\times10^{48}$ & (1.0$\pm$0.2)$\times10^{6}$  \\
6&G32.587+0.006  & 5.3 &5834$\pm$230  & 0.9 & (3.5$\pm$0.3)$\times10^{48}$ & (6.6$\pm$0.1)$\times10^{5}$  \\
7&G32.976-0.334  & 6.1 &6079$\pm$246  & 0.9 & (7.9$\pm$0.8)$\times10^{48}$ & (1.1$\pm$0.1)$\times10^{6}$   \\
8&G33.718-0.410  & 6.5 &6079$\pm$246  & 0.9 & (2.2$\pm$0.5)$\times10^{47}$ & (3.0$\pm$0.2)$\times10^{5}$  \\
9&G34.325+0.211  & 5.7 &6211$\pm$254  & 0.9 & (4.3$\pm$0.5)$\times10^{48}$ & (9.6$\pm$0.3)$\times10^{5}$  \\
10&G38.353-0.134 & 5.8 &5958$\pm$238  & 0.9 & (7.6$\pm$0.6)$\times10^{47}$ & (6.5$\pm$0.1)$\times10^{5}$   \\
11&G38.643-0.227 & 5.9 &5997$\pm$241  & 0.9 & (8.7$\pm$0.2)$\times10^{47}$ & (6.0$\pm$0.4)$\times10^{5}$  \\
12&G38.738-0.140 & 6.6 &6004$\pm$241  & 0.9 & (2.2$\pm$0.2)$\times10^{48}$ & (4.9$\pm$0.1)$\times10^{5}$  \\
13&G38.930-0.386 & 6.0 &6240$\pm$256  & 0.9 & (3.6$\pm$1.7)$\times10^{46}$ & (1.1$\pm$0.2)$\times10^{6}$  \\
14&G41.132-0.558 & 6.1 &6055$\pm$244  & 0.9 & (6.4$\pm$1.1)$\times10^{46}$ & (8.4$\pm$0.4)$\times10^{6}$   \\
15&G41.239-0.176 & 7.6 &6076$\pm$246  & 0.9 & (4.4$\pm$0.9)$\times10^{48}$ & (5.2$\pm$0.2)$\times10^{5}$  \\
16&G41.928+0.029 & 6.3 &6551$\pm$276  & 0.9 & (1.0$\pm$0.6)$\times10^{49}$ & (7.3$\pm$0.2)$\times10^{5}$  \\
17&G43.738+0.114 & 6.1 &6082$\pm$246  & 0.9 & (6.3$\pm$1.0)$\times10^{46}$ & (1.7$\pm$0.3)$\times10^{6}$   \\
18&G43.770+0.070 & 6.1 &6031$\pm$243  & 0.9 & (1.8$\pm$0.3)$\times10^{48}$ & (6.7$\pm$0.3)$\times10^{5}$   \\
19&G44.339-0.827 & 6.1 &6090$\pm$247  & 0.9 & (6.7$\pm$1.8)$\times10^{47}$ & (9.3$\pm$0.6)$\times10^{5}$  \\
20&G45.770-0.372 & 6.3 &6138$\pm$250  & 0.9 & (5.5$\pm$0.7)$\times10^{46}$ &  (7.4$\pm$0.8)$\times10^{5}$  \\
21&G46.176+0.536 & 7.9 &6642$\pm$282  & 0.9 & (3.7$\pm$1.1)$\times10^{47}$ & (5.7$\pm$0.2)$\times10^{5}$    \\
22&G47.028+0.232 & 6.3 &6155$\pm$251  & 0.9 & (1.7$\pm$0.2)$\times10^{48}$ & (6.6$\pm$0.1)$\times10^{5}$  \\
23&G49.738-0.616 & 6.1 &6090$\pm$247  & 0.9 & (1.2$\pm$1.2)$\times10^{47}$ & (1.3$\pm$0.5)$\times10^{6}$\\
24&G50.039-0.274 & 6.3 &6144$\pm$250  & 0.9 & --- & ---   \\
25&G50.079+0.571 & 8.5 &6839$\pm$294  & 0.9 & (3.0$\pm$0.5)$\times10^{48}$ & (6.4$\pm$0.3)$\times10^{5}$  \\
\hline\hline
\end{tabular}
\end{center}
\end{table*}

\subsection{CO molecular gas associated with  H {\small II} regions}
\subsubsection{The morphology of the molecular clouds}
To show the morphology of the molecular clouds associated with H {\small II} regions, we made the velocity-integrated intensity maps of $^{13}$CO(1-0) superimposed on each H {\small II} region map.  The velocity ranges are obtained from the spectrum marked by two vertical dashed lines in Figure 1 (Left panels). In Figure 1 (Right panels), each H {\small II} region may coincide with the surrounding molecular clouds. Based on the morphology of molecular clouds associated with H {\small II} region, these H {\small II} regions will be divided into three groups:

1. Having compact molecular cores are regularly spaced on borders of the H {\small II} region traced by PAHs emission, such as G19.504-0.193, G19.813+0.010, G23.029-0.405, G26.824+0.380, G32.587+0.006, G32.976-0.334, G33.718-0.410, G38.643-0.227, G38.738-0.140, G38.930-0.380, G41.928+0.029, G43.738+0.114, G43.770+0.070, G46.176+0.536, and G50.039-0.274. Massive stars form in giant molecular clouds. When UV radiation from these stars products H {\small II} regions, they may not far away from their natal molecular clouds. With the expansion of the H {\small II} regions, some dense cores may form in a compressed layer of molecular clouds between the ionization front and shock front preceding it in the molecular clouds. Gradually, dense core retain in the compressed layer, while the low density of molecular clouds may be dispersed.

2. Some dense clumps form between H {\small II} regions and the surrounding giant molecular clouds, such as G29.007+0.076, G34.325+0.211, G38.353-0.134, G41.132-0.558, G41.239-0.176, G44.339-0.827, G45.770-0.372, G47.028+0.232, and G49.738-0.616. When massive stars form on or near borders of giant molecular clouds, the expansion of H {\small II} regions accumulate the molecular gas between the giant molecular clouds and H {\small II} region. Hence, some dense clumps form in the compressed layer.

3. Molecular clouds cover the whole H {\small II} regions, such as G50.079+0.571. It indicates that these H {\small II} regions are still embedded in the giant molecular clouds.

In H {\small II} regions G29.007+0.076, G44.339-0.827, and G47.028+0.232, the PAHs emission all show a cometary structure. Moreover, the whole filamentary molecular cloud interacted with PAHs emission of G47.028+0.232 exhibits a cometary structure. The three H {\small II} regions support the RDI model. The rest of the H {\small II} regions in the groups 1 and 2, the molecular clouds with compact core regularly distribute around the PAHs emission, it is strong evidence in favor of the CC model. In addition, with the evolution of H {\small II} region, the H {\small II} regions in group 3 may become the morphology of that in groups 1 and 2.

\subsubsection{The physical parameters of the molecular clouds}
Assuming local thermodynamical equilibrium (LTE) and using the
$^{13}$CO(1-0) line,  the column density of the clumps in
cm$^{-2}$ was estimated to be \cite{Garden}
\begin{equation}
\mathit{N_{\rm H_{2}}}=1.25\times10^{15}\int T_{\rm mb}\rm dv ~~cm^{-2},
\end{equation}\indent
where $\rm dv$ is the velocity range in km s$^{-1}$. Here we
adopt a relation $N_{\rm H_{2}}$ $\approx$
$5\times10^{5}N_{\rm ^{13}CO}$ (\cite{Simon01}) and the average excitation temperature of 20 K from previous studies (\cite{Paron11,Dirienzo12}). If the clumps are approximately spherical in shape, the mean number density $\rm
H_{2}$ is $n(\rm i)$=$1.62\times10^{-19}N_{\rm H_{2}}/L$, where
$L$ is the clump diameter in parsecs (pc). The clump may be irregularly shaped, we use an effective radius, which is the radius of a circle that has the same solid angle on the sky as the clump. The derived column density of each clump is listed in Table 1, which range from 2.3$\times10^{21}$ cm$^{-2}$ to 7.3$\times10^{22}$ cm$^{-2}$. These values are compatible with the theoretical predictions of \cite{Whitworth94}, who show that for a wide range of input parameters, the gravitational fragmentation of a shocked layer occurs when the column density of this layer reaches a value $\sim$6.0$\times10^{21}$ cm$^{-2}$. In 25 H {\small II} regions, the derived column density of 21 H {\small II} regions are lower than value of \cite{Whitworth94}, implying that the shocked layer of these H {\small II} regions have been gravitationally fragmented.

\begin{figure}
\vspace{0mm}
\centering
\includegraphics[angle=0,scale=0.75]{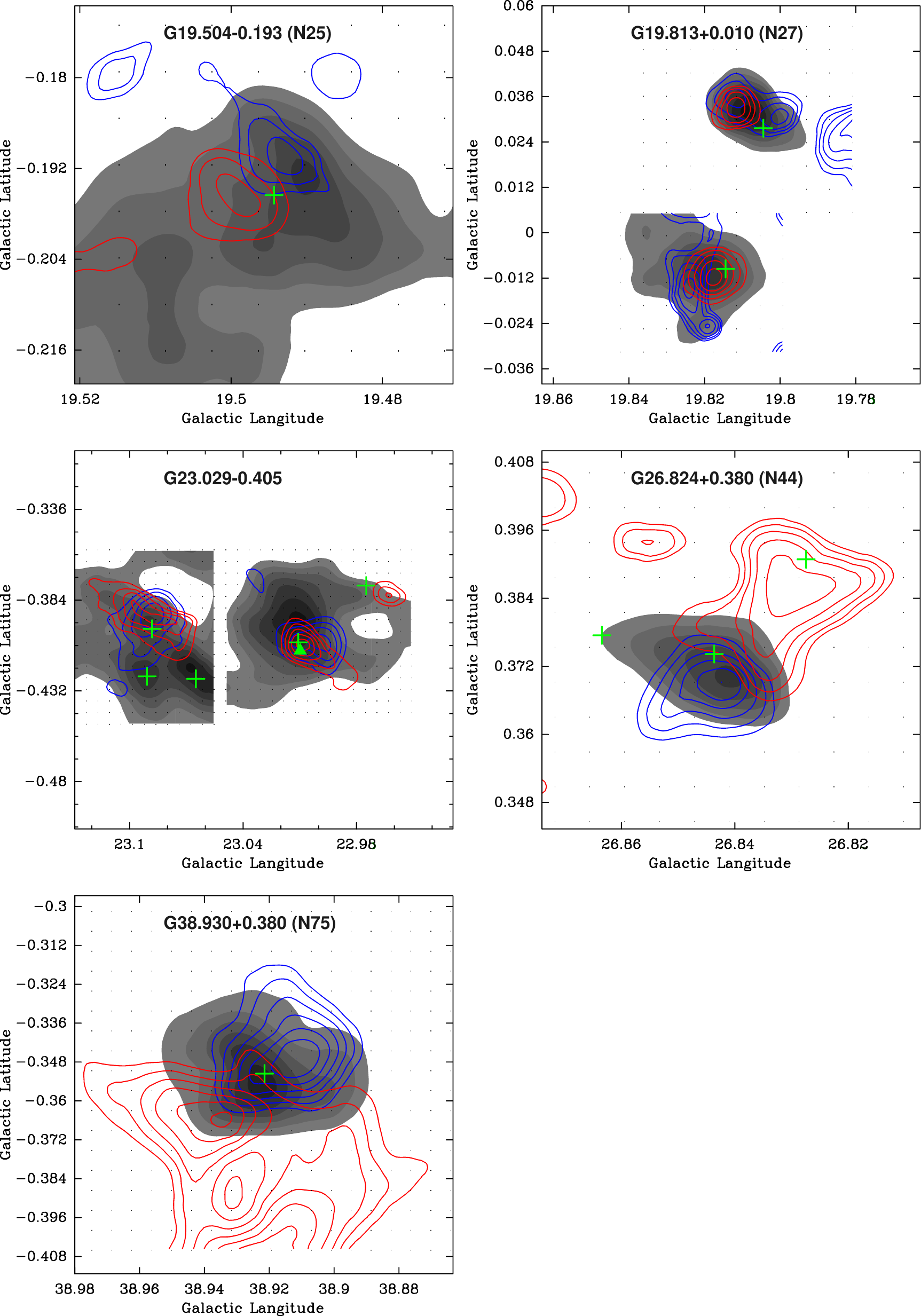}
\vspace{0mm}\caption{The contour maps of Outflow overlaid with the
$^{13}$CO(1-0) emission of each clump (gray scale). G19.504-0.193: The blue contour levels are 0.9, 1.0, 1.1 km s$^{-1}$, while the red contour levels are 0.9, 1.0, 1.1 km s$^{-1}$. G19.813+0.01: (Top) The blue contour levels are 0.5, 0.6, 0.7, 0.8 km s$^{-1}$, while the red contour levels are 0.5, 0.6, 0.7, 0.8 km s$^{-1}$. (Bottom) The blue contour levels are 0.2, 0.3, 0.4, 0.5, 0.6 km s$^{-1}$, while the red contour levels are 0.9, 1.1, 1.2, 1.3, 1.4 km s$^{-1}$. G23.029-0.405: (Left) The blue contour levels are 10.0, 10.7, 7.0, 8.0, 9.0, 10.0 km s$^{-1}$, while the red contour levels are 5.0, 6.0, 7.0, 8.0 km s$^{-1}$. (Right) The blue contour levels are 10.0, 10.7, 11.4, 12.1, 12.8, 13.6, 14.3 km s$^{-1}$, while the red contour levels are 10.0, 10.7, 11.4, 12.1, 12.8, 13.6, 14.3 km s$^{-1}$.  G26.824+0.380: The blue contour levels are 0.5, 0.6, 0.7, 0.8, 0.9 km s$^{-1}$, while the red contour levels are 0.4, 0.5, 0.6, 0.7 km s$^{-1}$.  G38.930+0.380: The blue contour levels are 2.4, 3.1, 3.8, 4.5, 5.2, 6.0, 6.7 km s$^{-1}$, while the red contour levels are 1.8, 2.3, 2.8, 3.3, 3.8, 4.3, 4.8 km s$^{-1}$.  }
\end{figure}

\subsubsection{Outflow detected}
Millimeter-wavelength continuum surveys of the Galactic
plane provide the most efficient way to find molecular clumps
that are the likely formation sites of massive stars and star
clusters. According to the Bolocam Galactic Plane Survey (BGPS), we selected the BGPS sources in the surrounding of each H {\small II} region. From Figure 1 (Right), we can see that there are several BGPS sources (green plus) distributed along the bubble of each H {\small II} region, except for G41.132-0.558, G41.928+0.029, G43.738, G43.770, G44.339-0.827, G46.176+0.536, G49.738-0.616, G50.039-0.274 and G50.079+0.571. Furthermore, some BGPS sources are located at the center of molecular cores. An outflow is strong evidence of the earlier star-forming activity. In order to search for the activity of star-forming activity, we detect outflows for these BGPS sources associated with H {\small II} regions and molecular cores. If several cores are linked together, then it is difficult to determine the component of the outflows. Hence, we only selected the BGPS sources  associated with the isolated cores. In 25 H {\small II} regions, we find that the molecular cores associated with 7 H {\small II} regions satisfy the above criterion. For these molecular cores, we first drew position-velocity (P-V) diagrams. According to the P-V diagrams, we
selected the integrated range of wings and determined the outflow
intensities of red and blue lobes. Seven molecular outflows
were identified by the contours of integrated intensities of $^{13}$CO(1-0) line wings in H {\small II} regions G19.504-0.193, G19.813+0.01, G23.029-0.405, G26.824+0.380 and G38.930+0.380. For the H {\small II} region G38.930+0.380,  we have performed a detailed analysis with $^{12}$CO $J$=1-0, $^{13}$CO $J$=1-0 and C$^{18}$O $J$=1-0 lines (\cite{Xu13}). The dynamic timescale of each outflow is estimated by equation $t_{_{\rm {\small out}}}$=9.78$\times$ $10^{5}$R/V (yr), where V in km $\rm s^{-1}$ is the maximum flow velocity relative to the cloud systemic velocity, and R in pc is the outflow size defined
by the length of the begin-to-end flow extension for each blueshifted and redshifted lobes. The mean dynamic timescale of each outflow is listed in Table 3.

\begin{table*}
\begin{center}
\tabcolsep 2.4mm\caption{The obtained parameters of outflow
}
\begin{tabular}{cccccccccccccccc}
\hline\hline
 Number &Region  &BGPS & $l$ & $d$ &  $t_{_{\rm {\small out}}}$ & \\
 &      &   &  deg  &   deg  & yr &\\
\hline
1&G19.504-0.193  & G19.490-0.197 &19.494  & -0.196 & 2.7$\times10^{5}$ \\
2&G19.813+0.010  & G19.818-0.009 &19.814  & -0.010 & 6.5$\times10^{5}$  \\
 &               & G19.806+0.033 &19.804  & 0.028  & 6.4$\times10^{5}$ \\
3&G23.029-0.405  & G23.012-0.410 &23.011  & -0.406 & 2.0$\times10^{5}$ \\
 &               & G23.090-0.394 &23.088  & -0.399 & 2.5$\times10^{5}$ \\
4&G26.824+0.380  & G26.843+0.375 &26.844  & 0.374  & 3.8$\times10^{5}$  \\
5&G38.930+0.380  & G38.920-0.352 &38.921  & -0.352 & 5.1$\times10^{5}$   \\
\hline\hline
\end{tabular}
\end{center}
\end{table*}

\section{Discussion}
\subsection{Assessment of triggered star formation}
$IRAS$ Point Source Catalog has reveal a number of stellar objects, which are younger than pre-main sequence stars (\cite{Beichman86,Fukui89}). Hence, the $IRAS$ point source is a good signpost of very early stages of star formation.
In order to explore a causal relationship between the H {\small II} regions and star
formation, we have searched for protostellar candidates in the
$IRAS$ Point Source Catalog that fulfill the following selection
criteria (Jankes et al. 1992 \& Xu et al. 2011): (1) $F_{100}\geq20$ Jy, (2) $1.2\leq
F_{100}/F_{60}\leq6.0$, (3) $F_{25}\leq F_{60}$, and (4) $R_{
IRAS}\leq 1.5\cdot R_{\rm H {\small II}}$, where $ F_{25}, F_{60}$ and
$F_{100}$ are the infrared fluxes at three IR bands ($25\mu \rm m,
60\mu \rm m$ and $100\mu \rm m$), respectively. The first criterion
selects only strong sources. The second and third discriminate
against cold IR Point sources probably associated with cool stars,
planetary nebulae, and cirrus clumps. While the fourth guarantees
that the search diameter ($R_{
IRAS}$)  include the complete surface of H {\small II} regions. $IRAS$ sources were found in
a search circle within 1.5 radius of $R_{\rm H {\small II}}$ centered at each H {\small II} region. In 19 H {\small II} regions, we find 23 young $IRAS$ source satisfied above criteria. The name,
coordinates, and flux at the 12, 25, 60, 100 $\mu$m  of these $IRAS$ point sources are listed in Table 3. Infrared luminosity (\cite{Casoli86}) and
dust temperature (\cite{Henning90}) are expressed respectively
as,
\begin{eqnarray}
\mathit{L}_{\rm IR}=(20.653\times F_{12}+7.538\times F_{25}+4.578\times F_{60}+1.762
\times F_{100})\times D^{2}\times0.30,
\end{eqnarray}
\begin{equation}
T_{\rm d}=\frac{96}{(3+\beta)\ln(100/60)-\ln(F_{60}/F_{100})},
\end{equation}

Where $D$ is the distance from the sun in kpc. The emissivity index
of dust particle ($\beta$) is assumed to be 2. The calculated
results are presented in Table 3. From the Table 3, we can see that the infrared luminosity $L_{\rm IR}$ of each $IRAS$ source is larger than 10$^{3}$$L_{\odot}$, which is associated with that of  bubble N131 (\cite{Zhang13}). The $IRAS$ sources associated with the BRCs have $L_{\rm IR}$ from $\sim$10 to 10$^{3}$$L_{\odot}$ (\cite{Sugitani91}). In addition, \cite{Junkes92} found 17 $IRAS$ sources associated with SNR G54.4-0.3, 16 of these sources $L_{\rm IR}$ are $\leq$ 10$^{3}$$L_{\odot}$. Similarly, the luminosity $L_{\rm IR}$ of all the $IRAS$ sources surrounding SNR IC443 are $<$ 10$^{3}$$L_{\odot}$ (\cite{Xu11}). BRCs found in H {\small II} regions are potential sites of triggered star formation by RDI process. We suggest that the process of star formation triggered by SNRs may be similar to RDI. Figure 1 (Right panels) also shows that the spatial
distribution of the selected $IRAS$ sources in 17 H {\small II} regions. From Figure
1 (Right panels), we see that all the $IRAS$ sources are associated with PAH emission and some $IRAS$ sources are located at the peak of $^{13}$CO molecular clumps, indicateing that these $IRAS$ sources with the larger infrared luminosity  may be triggered by CC process.
\begin{table*}
\begin{center}
\small \tabcolsep 2mm\caption{Selected IR point sources near 25
H {\small II} regions: IR flux densities, dust temperatures and IR
luminosities.}
\begin{tabular}{clccccccccccccr}
\hline\hline  Number & Source & l & b & $F_{12}$ & $F_{25}$
& $F_{60}$ & $F_{100}$ & $T_{\rm d}$ & $L_{\rm IR}$  \\
 & & deg &deg& [Jy] & [Jy]& [Jy]& [Jy]& [K]&[$\times10^{4}$$L_{\odot}$]\\
 \hline
1  & IRAS 18244-1201 &19.524 &-0.185 & 5.06&33.3 &732.7&2011.0&26.9 &35.7 \\
2 & - &- &- & -&-&- &- &-&-  \\
3 & IRAS 18317-0903 &22.987 &-0.381 &4.8  &45.1 &1144.0 &4999.0  &23.8 &9.6 \\
  & IRAS 18319-0903 &23.012 &-0.422 &4.8  &25.0 &1144.0 &4592.0  &24.3 &9.1 \\
4 & - &- &- & -&-&- &- &-&-  \\
5 & IRAS 18413-0331 &28.995 &0.082  &4.8  &14.3 &335.1  &577.5   &31.0 &11.0  \\
6 & IRAS 18482-0021 &32.590 &0.001  &10.0 &8.2  &109.7  &470.8   &23.9 &4.5 \\
7 & IRAS 18501-0009 &32.998 &-0.338 &3.0  &10.4 &162.1  &737.1   &23.5 &7.9 \\
8 & IRAS 18518+0026 &33.716 &-0.420 &3.1  &1.0  &39.8   &224.7   &22.4 &0.2  \\
9 &IRAS 18507+0118 &34.362 &0.206  &2.5  &12.1 &765.4   &1948.0  &27.5 &22.5 \\
10 & - &- &- & -&-&- &- &-&- \\
11 &IRAS 19002+0454 &38.645 &-0.225 &2.2  &11.3 &120.3   &285.6 &28.1 &2.7 \\
12 & -- &-- &-- & --&--&-- &-- &--&--  \\
13 &IRAS 19015+0503 &38.933 &-0.456 &2.2  &6.8  &497.2    &1156.0 &28.2 &14.6  \\
14 &IRAS 19060+0657 &41.136 &-0.553 &3.2  &2.8  &55.1     &141.8 &27.4 &0.7 \\
15 & - &- &- & -&-&- &- &-&- \\
16 &IRAS 19054+0754 &41.916 &0.007 &1.6  &1.5  &16.8     &51.0 &26.2 &0.8 \\
   &IRAS 19054+0755 &41.923 &0.035 &0.6  &4.1  &84.4     &186.7 &28.7 &3.0 \\
17 &IRAS 19086+0935   &   43.762 &   0.108  &   2.1 & 1.5 & 73.8& 356.7& 2.3  & 1.1  \\
18 &IRAS 19088+0935   &   43.788 &   0.059  &   3.1 & 17.1& 71.0&235.1& 25.6  & 1.1  \\
19 &IRAS 19130+0940   &   44.356 &   -0.822 &   3.5 & 0.9 & 150.9&557.2& 24.9  & 2.0  \\
20 & - &- &- & -&-&- &- &-&-  \\
21 &IRAS 19116+1156   &   46.195&    0.539 &   7.1 & 6.6 & 110.2 &391.7& 25.1  & 4.9  \\
   &IRAS 19116+1157   &   46.213&    0.548 &   3.9 & 3.2 & 388.3 &1297.0& 25.5  & 14.6  \\
   &IRAS 19116+1155   &   46.182&    0.531 &   2.5 & 5.2 & 486.2 &1454.0& 26.3  & 17.1  \\
22 &IRAS 19143+1232   &   47.029&    0.241 &   3.0 & 6.9 & 169.2 &519.4& 26.1  & 1.8  \\
23 &IRAS 19227+1431   &   49.746&   -0.623 &   3.1 & 2.2 & 24.8  &75.7 & 26.2  & 0.3  \\
24 & - &- &- & -&-&- &- &-&-  \\
25 &IRAS 19191+1522   &   50.075&   -0.560 &   2.5 & 7.6 & 206.4  &607.1 & 26.4  & 7.8  \\
\hline
\end{tabular}
\end{center}
\end{table*}

To further search for primary tracers of the star-formation activity in the
surroundings of H {\small II} regions, we used the GLIMPSE I catalog, which
consists of point sources that are detected at least twice in one
band.  Based on the photometric criteria of
\cite{Allen04}, we search for young star objects (YSO) candidates within a circle
of $>$1.5 radius centered at each H {\small II} regions. From the database, we constructed a $[5.8]-[8.0]$ versus
$[3.6]-[4.5]$ color-color (CC) diagram for each H {\small II} region to identify Class I
and Class II YSOs. Class I sources (10$^{5}$ yr) are protostars
with circumstellar envelopes, Class II sources (10$^{6}$ yr) are
disk-dominated objects. Figure 1 (Right panels) shows the spatial distribution of both Class I and Class
II sources. From Figures 1 (Right panels), we note that some Class I
sources are asymmetrically distributed only in H {\small II}
region G29.007+0.076, G44.339-0.827, and G47.028+0.232, and are mostly concentrated on the interacting regions between these H {\small II} regions and the surrounding molecular clouds. In H {\small II} regions G29.007+0.076 and G47.028+0.232, the Class I sources in the interacting regions are closed to the center of  the H {\small II} regions compared to the Class II sources. The existence of Class I and Class II sources may also indicate star formation activity.

Comparing the age of each H {\small II} region with the
characteristic star-formation timescales, we conclude that the three  H {\small II} regions can trigger the clustered star formation. In addition, we find that the PAHs emission of the three H {\small II} regions is stronger closed to the molecular clouds, and all show the cometary globule, suggesting these Class I and Class II sources may be triggered by RDI process. In addition, we for the first time detected seven molecular outflows in H {\small II} regions G19.504-0.193, G19.813+0.01, G23.029-0.405, G26.824+0.380 and G38.930+0.380.  The mean dynamic timescale of each outflow is larger than the age of the corresponding H {\small II} regions. Hence, we conclude that these outflow sources may be triggered by  the corresponding H {\small II} regions, but the observations at higher spatial resolution are needed to resolve the detailed kinematics of the outflows in these H {\small II} regions.

\section{Conclusions}
\label{sect:conclusion}
In this work, we have performed a detailed analysis of the environment of 25 H {\small II} regions with bubble morphologies in $^{13}$CO(1-0) and infrared data. These H {\small II} regions at 8 $\mu$m show the morphology of the complete and partial bubble, indicating PAHs emission is destroyed inside the ionized region. Each H II region may be associated with the surrounding molecular clouds. In 25 H {\small II} regions, the derived column density of 21 H {\small II} regions are larger than value of Whitworth et al. (1994), implying that the shocked layer of these H {\small II} regions have been gravitationally fragmented. We derived that the electron temperature ranges from 5627 K to 6839 K in these H {\small II} regions, and the averaged electron temperature is 6083 K. Moreover, the age of all the H {\small II} regions is from 3.0$\times10^{5}$ yr to 1.7$\times10^{6}$ yr, and the mean age is 7.7$\times10^{5}$ yr.  We selected 23 young $IRAS$ sources with the larger infrared luminosity ($>$10$^{3}$$L_{\odot}$) in 19 H {\small II} regions, which may be triggered by CC process. In addition, some young stellar objects (YSOs) (including Class I sources) are concentrated around H {\small II}
regions G29.007+0.076, G44.339-0.827, and G47.028+0.232, which appear to be sites of
ongoing star formation. The PAHs emission of the three H {\small II} regions all show the cometary globule. Comparing the
age of each H {\small II} region and the characteristic timescales for star formation, we
conclude that the three H {\small II} region can trigger the clustered star formation by RDI process. In addition, we for the first time detected seven molecular outflows in the five H {\small II} regions. These outflow sources may be related to the corresponding H {\small II} regions.

\begin{acknowledgements}
We thank an anonymous referee for very useful suggestions
\end{acknowledgements}

\label{lastpage}

\end{document}